\documentclass[12pt]{elsarticle}
\usepackage{amsmath,amssymb,comment,color,braket,epsfig,graphicx,hyperref,here,natbib,slashed,subfigure,mathrsfs}
\graphicspath{{./Figures/}}
\def\nn{\nonumber}
\def\D{\mathrm{d}}
\def\beq{\begin{equation}}
\def\eeq{\end{equation}}
\def\bea{\begin{eqnarray}}
\def\eea{\end{eqnarray}}
\begin{document}
%==================================
\begin{frontmatter}
\title{Revisiting quark-hadron duality for heavy meson non-leptonic decays in two-dimensional QCD}
\author{Hiroyuki Umeeda}
\ead{umeeda@gate.sinica.edu.tw}
\address{Institute of Physics, Academia Sinica, Taipei 11529, Taiwan, Republic of China}
\begin{abstract}
We study lifetimes of heavy mesons in the 't Hooft model, a large-$N_c$ theory of strong interaction in two-dimensional spacetime. Since this model is solvable, one can evaluate the total decay widths through hadronic amplitudes that are determined unambiguously within the formalism. We investigate quark-hadron duality by numerically comparing the exclusive result and one from the heavy quark expansion, with the contribution of Pauli interference being consistently incorporated. Within certain numerical accuracy, we find that the maximal difference between inclusive and exclusive rates for $\tau[D^+_s]/\tau[D^0]$ $(\tau[B^+]/\tau[B^0])$ is as large as the current experimental (theoretical) error in four-dimensions unlike $\tau[D^+]/\tau[D^0]$, where an excellent agreement between the rates is seen.
\end{abstract}
\end{frontmatter}
%=======================
\section{Introduction}\noindent
%=======================
In investigating heavy flavor physics, inclusive processes play a certain important role. Decay rates for those processes are analyzed on the basis of Wilson's operator product expansion (OPE) \cite{Wilson:1969zs}, which is adapted \cite{Shifman:1978bx} to quantum chromodynamics (QCD). Through analytic continuation to Minkowski domain, observables are expanded by an inverse of heavy quark mass, and the method is called the Heavy Quark Expansion (HQE) (see Refs.~\cite{Bigi:1997fj, Lenz:2014jha} for reviews and references therein). That is, provided that quark is sufficiently heavy, the HQE works as a promising tool to deal with heavy quarks. For the beauty hadrons, the lifetime ratios are well-described by HQE, as shown in the recent analyses of the HQE \cite{Kirk:2017juj, Cheng:2018rkz}. In addition, the width difference in the $B_s^0-\bar{B^0_s}$ mixing is in good agreement \cite{Kirk:2017juj, King:2019lal} with the experimental data of HFLAV \cite{HFLAV:2019otj} while the experimental precision is still not under control for the $B_d^0-\bar{B^0_d}$ mixing.
\par
Meanwhile, for the charm sector, the situation is rather subtle: since the mass scale of charm quark might not be heavy enough, the convergence of the HQE is often questioned. The difficulty is easily seen from the fact that the approximation treating charmed hadron decays as partonic processes \cite{Gaillard:1974mw, Ellis:1975hr, Cabibbo:1977zv} leads to identical lifetimes (see also early works \cite{Koide:1979iw, Sawayanagi:1982ms}), immediately failing to reproduce the experimental data \cite{HFLAV:2019otj} (see also the recent lifetime data for $D^0$ and $D^+$ from the Belle II experiment \cite{Belle-II:2021cxx}). As for the $D^0-\bar{D^0}$ mixing, due to the Glashow-Iliopoulos-Maiani mechanism \cite{Glashow:1970gm}, the leading contribution undergoes a strong cancellation so that a tiny error is possibly enlarged after the cancellation happens. It is shown that the HQE results \cite{Golowich:2005pt, Bobrowski:2010xg} which neglect higher dimensional contributions such as 6-quark and 8-quark operators are smaller than the experimental data \cite{HFLAV:2019otj} by $\mathcal{O}(10^{-4})$. For this issue, see a recent work \cite{Li:2020xrz} based on the dispersive approach indicating that the inclusive result is possibly enhanced.
\par
Another non-trivial aspect is whether quark-hadron duality \cite{Bloom:1970xb, Poggio:1975af} (see Ref.~\cite{Shifman:2000jv} for a review), a tacit assumption in HQE, is realized for $b$ and $c$ quarks. As methods to investigate this issue, the instanton-based approach \cite{Chay:1994si, Falk:1995yc, Chibisov:1996wf} and the resonance-based approach 
\cite{Shifman:1994yf,
Zhitnitsky:1995qa,
Colangelo:1997ni,
Grinstein:1997xk,
Blok:1997hs,
Bigi:1998kc,
Grinstein:2,
Bigi:1999fi,
Burkardt:2000ez,
Lebed:1,
Beane:2001uj,
Grinstein:2001zq,
Grinstein:2001nu,
Mondejar:2006ct,
Mondejar:2008pi,
Umeeda:2021llf} are considered (see also Ref.~\cite{Blok:1996fg} and Ref.~\cite{Blok:1997yd}, where duality in non-leptonic $B$ decay and QCD sum rules is investigated, respectively). The latter is particularly facilitated in the 't Hooft model \cite{tHooft:1974pnl}, a two-dimensional model of QCD together with large-$N_c$ limit. This model is a solvable theory, and accommodates the linear Regge trajectory for the squared meson mass spectrum, that plays a certain central role in the resonance-based approach. In this way, the theory offers a tractable method to study duality and its violation in heavy quark physics \cite{
Colangelo:1997ni,
Grinstein:1997xk,
Bigi:1998kc,
Grinstein:2,
Bigi:1999fi,
Burkardt:2000ez,
Lebed:1,
Grinstein:2001zq,
Mondejar:2006ct,
Umeeda:2021llf}.
\par
Recent HQE analyses for charmed baryons and/or mesons are carried out in Refs.~\cite{Lenz:2013aua, Kirk:2017juj, Cheng:2018rkz, King:2021xqp}. From those results, on one hand, we find that a subleading power correction, especially Pauli interference (PI) \cite{Guberina:1979xw}, gives a huge deviation of $\tau[D^+]/\tau[D^0]$ from unity. On the other hand,  it is also shown that the lifetime ratios for $D$ mesons are in agreement with the experimental data \cite{HFLAV:2019otj} under the theoretical uncertainties although the truncated perturbative QCD corrections might be non-negligible. While the concluding statements are different in the groups, the latest work \cite{King:2021xqp} claims that neither clear breakdown of the HQE nor violation of duality is observed for charmed meson lifetimes. However, given that non-zero uncertainties are involved in theoretical and experimental results, duality violation is not straightforwardly ruled out for heavy meson decays in principle. In this respect, care must be taken for the fact that duality violation may well be originated from the truncated power series \cite{Chibisov:1996wf, Shifman:2000jv} in the Euclidean OPE while the convergence of $1/m_c$ series is, at least, not good.
\par
In this letter, we study lifetimes or equivalently total widths of $D$ and $B$ mesons in the 't Hooft model, which is regarded as a laboratory to test QCD. By solving the Bethe-Salpeter equation \cite{Nambu:1950dpa, Salpeter:1951sz} in the light-cone gauge (the 't Hooft equation), one can obtain light-cone distribution functions as well as masses of mesons. Since hadronic decay amplitudes are calculable in this way, we can evaluate decay widths as a sum of exclusive processes in two-dimensions. By varying $m_b$ or $m_c$, we revisit quark-hadron duality via comparison between the inclusive width and sum of the exclusive ones. The Cabibbo-suppressed processes, giving $\mathcal{O}(5\%)$ corrections to the Cabibbo-favored rate are included in the analysis. Both charged-current and neutral-current, where the latter comes from the Fiertz rearrangement of the conventional operator basis, are included to analyze the color-allowed PI in a consistent manner. Through this method, we show how large difference between inclusive and exclusive for the lifetime ratio is possible in two-dimensions.
%=======================
\section{Formalism}\noindent
%=======================
We introduce the following effective Lagrangians yielding Cabibbo-allowed charmed meson decays,
\bea
\mathcal{L}_{|\Delta C|=1}&=&-\frac{G_F}{\sqrt{2}}V_{cs}^*V_{ud}[a_1^{(c)}(\bar{s}^\alpha \gamma_\mu c^\alpha)(\bar{u}^\beta\gamma^\mu d^\beta)+a_2^{(c)}(\bar{u}^\alpha \gamma_\mu c^\alpha)(\bar{s}^\beta\gamma^\mu d^\beta)]+\mathrm{H}.\mathrm{c}.,\qquad\quad
\label{Eq:Hamil}
\eea
where the sum over color indices are understood while $G_F$ is an overall normalization analogous to the Fermi constant in four-dimensions. However, the numerical results presented later are based on the ratios of widths, and are insensitive to $G_F$. In the above operator basis, the first term represents charged-current interaction while the second one gives the neutral current interaction. The effective Lagrangians inducing Cabibbo-suppressed processes, as well as the cases for the bottom quark, are also considered via proper replacements of flavor indices in Eq.~(\ref{Eq:Hamil}). \par
We introduced only the vector currents in Eq.~(\ref{Eq:Hamil}) as either vector or axial vector current is reducible in two-dimensions due to $\gamma_\mu=\epsilon_{\mu\nu}\gamma^\nu\gamma_5$. One can find that replacement of the vector currents by the axial vectors changes only the overall sign of the decay amplitude, while keeping the final results of total widths unchanged. In testing strong dynamics, it is preferable that parameters of the weak interaction, where quark-hadron duality is not relevant, are fixed to values similar to the realistic case. That is, non-trivial calculation lies only in the QCD side, while the weak interaction side is required to have a 4D-like strength. In view of this standpoint, we set parameters related with weak interactions to four-dimensional values in the large-$N_c$ limit. Specifically, the coefficients of the local operators are fixed by $a_1^{(c)}=1.20$ and $a_2^{(c)}=-0.39$ at the scale of $1.27~\mathrm{GeV}$ for charm \cite{King:2021xqp} and $a_1^{(b)}=1.08$ and $a_2^{(b)}=-0.19$ at the scale of $4.40~\mathrm{GeV}$ with $\Lambda_{\overline{\rm MS}}^{(5)}=225~\mathrm{MeV}$ and $m_t=170~\mathrm{GeV}$ for bottom \cite{Buras:1998raa}. The values of Cabibbo-Kobayashi-Maskawa matrix elements \cite{Cabibbo:1963yz,Kobayashi:1973fv} are extracted from PDG data \cite{ParticleDataGroup:2020ssz}. 
\par
Throughout this work, $Q$ denotes a heavy quark (either $c$ or $b$), while $\bar{q}$ represents a spectator anti-quark. We also introduce a notation of $H$ for a ground state pseudoscalar formed by $Q$ and $\bar{q}$. In the large-$N_c$ limit, the contributions of semi-leptonic decays are suppressed by $1/N_c$, relative to the leading term in $1/m_c$ expansion, and omitted below\footnote{For semi-leptonic decays, the agreement between inclusive and exclusive rates are numerically studied in Ref.~\cite{Lebed:1}, indicating a better agreement than non-leptonic decays \cite{Grinstein:1997xk,Grinstein:2}.}. 
The contributions of the leading non-leptonic decay as well as PI represented as an exclusive sum are given by,
\bea
\Gamma_{H,\:(q_1, q_2, q_3)}^{\rm (\mathbf{dec}, \: {\rm exc})}&=&a_1^2\displaystyle\sum_{k, m}\frac{\left|T^{(k, m)}_{(Q\bar{q}) (\bar{q}_1, q_2, q_3)}\right|^2}{4M_{H}^2|p_{km}|}+a_2^2\displaystyle\sum_{k, m}\frac{\left|T^{(k, m)}_{(Q\bar{q}) (\bar{q}_1, q_3, q_2)}\right|^2}{4M_{H}^2|p_{km}|},\label{Eq:LEADDEC}\\
\Gamma_{H,\:(q_1, q_2, q_3)}^{\rm (\mathbf{PI}, \: {\rm exc})}&=&a_1a_2\displaystyle\sum_{k, m}(-1)^{k+m+1}\frac{T^{(k, m)}_{(Q\bar{q}) (\bar{q}_1, q_2, q_3)}T^{(m, k)*}_{(Q\bar{q}_1) (\bar{q}, q_3, q_2)}}{2M_{H}^2|p_{km}|},\label{Eq:PAULI}
\eea
where $T^{(k, m)}_{(Q\bar{q})(\bar{q}_1, q_2, q_3)}$ is a color-allowed tree diagram defined in the following paragraph and $p_{km}$ is a momentum of a daughter meson in the rest frame of the initial meson. One finds a peculiar nature of the phase space in two-dimensions in Eqs.~(\ref{Eq:LEADDEC}, \ref{Eq:PAULI}), that is, proportionality to $|p_{km}|^{-1}$, instead of $|p_{km}|^{+1}$. Due to this factor, the hadronic width is enhanced when the heavy quark mass is slightly larger than a threshold value, as we shall see in some numerical result later.
\par
\begin{figure}[h]
 \begin{center}
  \includegraphics[width=40mm]{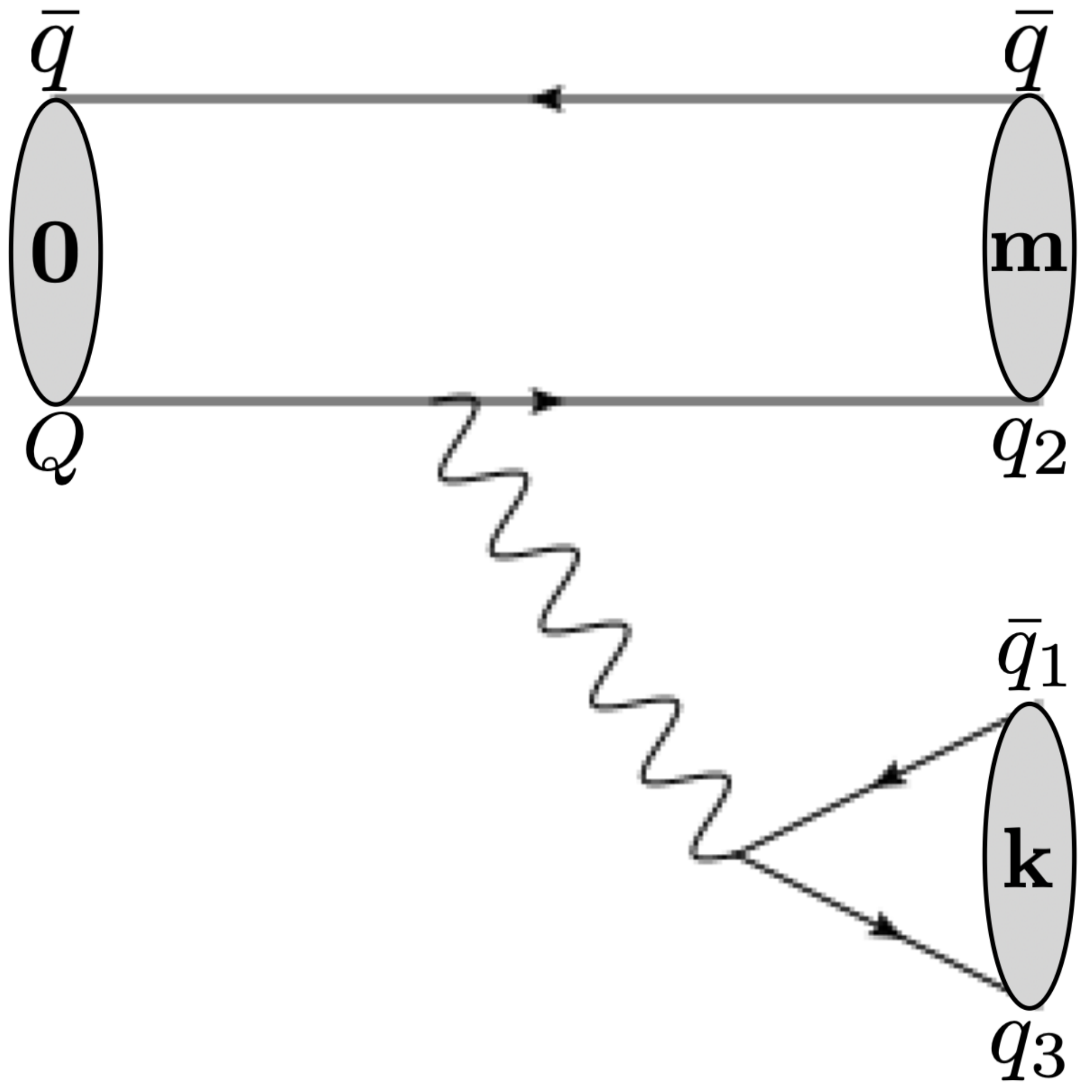}
 \end{center}
 \vspace{-5mm}
 \caption{Color-allowed tree diagram with index assignment. $\mathbf{0}, \mathbf{m}$ and $\mathbf{k}$ represent radial excitation numbers with $\mathbf{0}$ being a ground state.}
 \label{Fig:1}
\end{figure}

In order to ensure the numerical stability, the contribution of the triple overlap integral, giving the correction suppressed by at least $1/m_Q^2$ \cite{Bigi:1999fi}, is neglected in the numerical analysis as in the previous work \cite{Umeeda:2021llf}. In this way, we confirm that the numerical results are well-stabilized. Further improvement requires accurate evaluations of the overlap integrals with oscillating integrands. In the mentioned accuracy, the color-allowed tree diagram in Eqs.~(\ref{Eq:LEADDEC}, \ref{Eq:PAULI}) is given for on-shell mesons by \cite{Grinstein:1997xk,Bigi:1999fi},
\bea
T^{(k, m)}_{(Q\bar{q})(\bar{q}_1, q_2, q_3)}&=&\frac{G_F}{\sqrt{2}}\sqrt{\frac{N_c}{\pi}}c^{(k)}_{(q_3\bar{q}_1)}[(-1)^{k+1}M_k^2\mathcal{C}^{(k, m)}_{(Q\bar{q})(q_2)}+m_Qm_{q_2}\mathcal{D}^{(k, m)}_{(Q\bar{q})(q_2)}],\quad\label{Eq:topo}\\
\mathcal{C}^{(k, m)}_{(Q\bar{q})(q_2)}&=&-\frac{1-\omega^{(k, m)}}{\omega^{(k, m)}}\int_0^1\mathrm{d}x \phi^{(0)}_{(Q\bar{q})}[1-(1-\omega^{(k, m)})(1-x)]\phi^{(m)}_{(q_2\bar{q})}(x),\label{Eq:Cdou}\\
\mathcal{D}^{(k, m)}_{(Q\bar{q})(q_2)}&=&-\omega^{(k, m)}\int_0^1\mathrm{d}x
\frac{\phi^{(0)}_{(Q\bar{q})}[1-(1-\omega^{(k, m)})(1-x)]}{1-(1-\omega^{(k, m)})(1-x)}
\displaystyle\frac{\phi^{(m)}_{(q_2\bar{q})}(x)}{x},\label{Eq:Ddou}\\
\omega^{(k, m)}&=&\frac{1}{2}\left[1+\left(\frac{M_k^2-M_m^2}{M_0^2}\right)-\sqrt{1-2\left(\frac{M_k^2+M_m^2}{M_0^2}\right)+\left(\frac{M_k^2-M_m^2}{M_0^2}\right)^2}\right],\qquad
\eea
where $\phi^{(n)}_{Q\bar{q}}$ is a light-cone distribution function determined by solving the 't Hooft equation, that is explained in Sec.~\ref{Sec:3}, while $c^{(k)}_{(q_3\bar{q}_1)}$ is defined as a normalized meson decay constant given by the integral of $\phi^{(k)}_{(q_3\bar{q}_1)}$ from 0 to 1. In Eqs.~(\ref{Eq:topo}-\ref{Eq:Ddou}), $\mathcal{C}$ and $\mathcal{D}$ denote overlap integrals between the wave functions for the ground state heavy meson and the daughter meson. The indices of $T^{(k, m)}_{(Q\bar{q})(\bar{q}_1, q_2, q_3)}$ are concretely displayed in Fig.~\ref{Fig:1}.
%===================
%===================
\par
The inclusive rates to be contrasted with the exclusive ones in Eqs.~(\ref{Eq:LEADDEC}, \ref{Eq:PAULI}) are given by\footnote{In the massless limit of the light quarks (or $m_Q\to \infty$), duality between inclusive and exclusive rates can be confirmed analytically for PI \cite{Bigi:1999fi}.},
\bea
\Gamma_{H,\:(q_1, q_2, q_3)}^{(\rm \mathbf{dec}, \:{\rm inc})}&=&N_c\frac{G_F^2}{4\pi}\left(a_1^2\frac{m_Q^2-m_{q_2}^2}{m_Q}+a_2^2\frac{m_Q^2-m_{q_3}^2}{m_Q}\right)\frac{\bra{H}\bar{Q}Q\ket{H}}{2M_H},\label{Eq:decPar}\\
\Gamma_{H,\:(q_1, q_2, q_3)}^{(\rm \mathbf{PI}, \:{\rm inc})}&=&-a_1a_2G_F^2
\left[(F_{23}^{\rm (th)}+2G_{23}^{\rm (th)})\frac{\bra{H}(\bar{Q}\gamma_\mu\gamma_5 q)(\bar{q}\gamma^\mu\gamma_5 Q)\ket{H}}{2M_H}\right.\nn\\&&\left.+(G_{23}^{\rm (th)}+2H_{23}^{\rm (th)})\frac{\bra{H}(\bar{Q}i\gamma_5 q)(\bar{q}i\gamma_5 Q)\ket{H}}{2M_H}\right],\label{Eq:PIPar}
\eea
where Eq.~(\ref{Eq:decPar}) is extracted from Ref.~\cite{Bigi:1998kc}. Some functions that have non-zero values in physical regions are introduced in Eq.~(\ref{Eq:PIPar}) ($z_i = m_{q_i}^2/m_Q^2$),
\bea
F^{\rm(th)}_{ij}&=&\sqrt{1-2(z_i+z_j)+(z_i-z_j)^2},\\
G^{\rm(th)}_{ij}&=&\frac{z_i+z_j-(z_i-z_j)^2}{\sqrt{1-2(z_i+z_j)+(z_i-z_j)^2}},\\
H^{\rm(th)}_{ij}&=&\frac{\sqrt{z_iz_j}}{\sqrt{1-2(z_i+z_j)+(z_i-z_j)^2}},
\eea
In Eq.~(\ref{Eq:PIPar}) we do not explicitly write the contribution of the scalar bilinears to PI as it vanishes in the large-$N_c$ factorization. In four-dimensions, it is known that the phase space factor for PI is greater than that of the leading decay by 16$\pi^2$. This enhancement is absent in two-dimensions for the normalization of $|a_1|=|a_2|$ and $m_Q\to \infty$  as seen by the comparison between Eq.~(\ref{Eq:decPar}) and Eq.~(\ref{Eq:PIPar}). It should be also noted that the sign of Eqs.~(\ref{Eq:decPar}, \ref{Eq:PIPar}) are both positive with $a_1>0, a_2<0$, which is to be contrasted with the usual four-dimensional case, where PI gives negative contribution in heavy meson decays. Another point to be mentioned is PI gives $\mathcal{O}(1/m_{Q})$ corrections, relative to the leading decay operator. This is to be distinguished from the four-dimensional case, where PI gives $\mathcal{O}(1/m_{Q}^3)$ corrections. Combining these aspects, together with the CKM factors multiplied later in Eqs.~(\ref{Eq:dec_D}, \ref{Eq:PI_D}), one finds that $D^+$ has a shorter lifetime than $D^0$ in the two-dimensional case. 
\par
As for weak annihilation (WA) diagrams, it is broadly recognized in the literature (see {\it e.g.,} Refs.~\cite{Grinstein:1997xk, Bigi:1998kc, Bigi:1999fi}) that $\Gamma^{(\rm WA)}_H\propto N_c^2$ in the large-$N_c$ limit, to be contrasted with $\Gamma^{(\rm dec)}_H\propto N_c$ and $\Gamma^{(\rm PI)}_H\propto N_c$, so that WA is dominant in the counting of $N_c$, for the inclusive side. However, in the large-$N_c$ limit, it is immediately seen that the naive $1/N_c$ counting of the exclusive side is proportional to $N_c$, in contradiction to the inclusive side. Then, the exclusive side should be enhanced by $N_c$ through the resonant intermediate states with smearing \cite{Grinstein:2}. This procedure entails the analysis beyond the rigorous large-$N_c$ limit since otherwise the effect of the resonant width vanishes. Furthermore, an {\it ad-hoc} smearing function is required to discuss global duality. As the entirely consistent treatment of the $1/N_c$ correction gives certain modifications to the theoretical framework, the study of duality for combined PI and WA is beyond the scope of the current work. In the four-dimensional case, it is shown that the contribution of WA approximately vanishes in the vacuum insertion for the four-quark operator's matrix element, and its impact is smaller than PI for heavy meson decays \cite{Lenz:2014jha, Cheng:2018rkz, Lenz:2013aua, King:2021xqp}. We refer the reader to the dedicated work \cite{Grinstein:2} with the finite-$N_c$ to study quark-hadron duality for WA in the 't Hooft model. 
\par
For the matrix element of the leading decay operator in Eq.~(\ref{Eq:decPar}), we follow the argument obtained in Ref.~\cite{Bigi:1998kc}: by integrating 't Hooft equation with some weight, one can derive sum rules for the meson wave functions. Combining them, and using the completeness relation that the wave function of the daughter meson satisfies, we have another sum rule. An upshot of this argument \cite{Bigi:1998kc} leads to,
\bea
\frac{\bra{H}\bar{Q}Q\ket{H}}{2M_H}&=&\frac{m_Q}{M_H}\int_0^1\frac{\mathrm{d}x}{x}\phi_H^2(x)+\mathcal{O}\left(\frac{1}{m_Q^5}\right).\label{Eq:SHIF}
\eea
The use of Eq.~(\ref{Eq:SHIF}) enables us to go beyond the approximation of $\bra{H}\bar{Q}Q\ket{H}/2M_H\to 1$ similarly to the usual four-dimensional analysis, where the contributions of $\mu_\pi, \mu_G$ and $\rho_D$ are included as higher order terms. We adopt Eq.~(\ref{Eq:SHIF}) in the inclusive numerical result since it does not explicitly rely on exclusive final states.
\par
As for PI, the associated matrix elements are factorized in the large-$N_c$ limit,
\bea
\frac{\bra{H}(\bar{Q}\gamma_\mu\gamma_5 q)(\bar{q}\gamma^\mu\gamma_5 Q)\ket{H}}{2M_H}&=&\frac{N_c}{2\pi}c_H^2M_H,\label{Eq:4A}\\
\frac{\bra{H}(\bar{Q}i\gamma_5 q)(\bar{q}i\gamma_5 Q)\ket{H}}{2M_H}&=&\frac{N_c}{2\pi}c_H^2M_H\left(\frac{M_H}{m_Q+m_q}\right)^2\label{Eq:4P}.
\eea
Here we introduced a shorthand notation $(H)$ for the ground state that consists of $Q\bar{q}$. The matrix element for the product of two scalar bilinears vanishes since it does not preserve parity. In the large-$N_c$ limit, both Eq.~(\ref{Eq:4A}) and Eq.~(\ref{Eq:4P}) do not explicitly include $1/m_Q$ corrections although they are contained implicitly in $c_H$ and $M_H$ on r.h.s.
\par
For both inclusive and exclusive sides, the total widths including the CKM factors are written by ($\alpha={\rm inc}, {\rm exc}$),
\bea
\Gamma_{D_q}^{(\mathbf{dec},\:\alpha)}&=&\displaystyle\sum_{q_2=d, s}\displaystyle\sum_{q_1=d, s}|V_{cq_2}^*V_{uq_1}|^2\:\Gamma_{D_q,\: (q_1, q_2, u)}^{ (\mathbf{dec},\: \alpha)},\quad (q=u, d, s)\label{Eq:dec_D}\\
\Gamma_{D_q}^{(\mathbf{PI},\:\alpha)}&=&\displaystyle\sum_{q_2=d, s}|V_{cq_2}^*V_{uq}|^2\:\Gamma_{D_q,\: (q, q_2, u)}^{(\mathbf{PI},\: \alpha)},\quad (q=d, s)\label{Eq:PI_D}\\
\Gamma_{\bar{B}_q}^{(\mathbf{dec},\:\alpha)}&=&\displaystyle\sum_{q_3=d, s}\displaystyle\sum_{q_2=u, c}\displaystyle\sum_{q_1=u, c}|V_{q_2b}V_{q_1q_3}^*|^2\:\Gamma_{\bar{B}_q,\: (q_1, q_2, q_3) }^{ (\mathbf{dec},\: \alpha)},\quad (q=u, d, s)\qquad\\
\Gamma_{\bar{B}_q}^{(\mathbf{PI},\:\alpha)}&=&\displaystyle\sum_{q_3=d, s}\displaystyle\sum_{q_2=u, c}|V_{q_2 b}V_{qq_3}^*|^2\:\Gamma_{\bar{B}_q, (q, q_2, q_3)}^{(\mathbf{PI},\:\alpha)},\quad (q=u)
\eea
Due to the CKM structure in Eq.~(\ref{Eq:PI_D}), PI for $D^+$ is CKM-favored whereas that for $D_s^+$ is Cabibbo-suppressed.\par
For the decay width of $D_s^+$, we do not consider $D^+_s\to \tau^+\nu_\tau$ in either inclusive process or exclusive one. This setup is similar to the four-dimensional analyses \cite{Lenz:2013aua, Cheng:2018rkz, King:2021xqp}, where the contribution of $D_s^+\to \tau^+\nu_\tau$ to the total width is subtracted in the experimental data, or equivalently the exclusive side.\par
$\Gamma^{(\rm \mathbf{dec},\:\mathrm{inc})}$ in Eq.~(\ref{Eq:decPar}) is proportional to $m_Q$ for $m_Q\to \infty$. Provided that quark-hadron duality is valid, the exclusive counterpart in Eq.~(\ref{Eq:LEADDEC}) should have the same asymptotic behavior accordingly. This is numerically tested in Ref.~\cite{Grinstein:1997xk}. This confirmed proportionality to $m_Q$ in two-dimension is to be  distinguished from $\Gamma^{(\rm \mathbf{dec},\:\mathrm{inc})}\propto m^5_Q$ in four-dimensions. In what follows, we analyze lifetime ratios that consist of non-leptonic decays, instead of the decay width in order to cancel the overall heavy quark mass dependence. In the numerical results, the lifetime ratios are not expanded by $1/m_Q$, and are calculated through the total width that arise from the sum of $\Gamma^{(\mathbf{dec})}$ and $\Gamma^{(\mathbf{PI})}$ for charged mesons, and $\Gamma^{(\mathbf{dec})}$ for neutral mesons.
%=======================
\section{Numerical results}
%=======================
\label{Sec:3}
For the determination of wave functions and masses of mesons, the 't Hooft equation is introduced by,
\bea
M_n^2\phi_{(q_1\bar{q}_2)}^{(n)}(x)=\left(\frac{m_1^2-\beta^2}{x}+\frac{m_2^2-\beta^2}{1-x}\right)\phi^{(n)}_{(q_1\bar{q}_2)}-\beta^2\: \mathrm{Pr}\int_0^1\D y\frac{\phi^{(n)}_{(q_1\bar{q}_2)}(y)}{(x-y)^2}\label{Eq:tHooftEq},
\eea
where $\beta^2 =g^2N_c/(2\pi)$. As a numerical method to solve the above integral equation, we adopt the BSW-improved Multhopp technique \cite{Brower:1978wm}. The detail of the formalism is summarized in the previous work \cite{Umeeda:2021llf}, and is not repeated here. We fix the bare coupling of QCD so as to fit the string tension of four-dimensional QCD, leading to $\beta = 340~\mathrm{MeV}$ \cite{Burkardt:2000uu,Jia:2017uul}. In discretizing the 't Hooft equation via the BSW method, the dimension for eigenvectors is taken by 500. Heavier 200 excited states, which do not follow the linear Regge trajectory, of the total 500 states are truncated as in Ref.~\cite{Umeeda:2021llf}. In this way, parameters of mesons for both ground and excited states are numerically determined in principle.
\par
Although a two-dimensional bare mass is not directly related to masses in four-dimension, we fix the former by the renormalized masses in four-dimensions as reference values. The strange quark mass is set by $m_s/\beta=0.32$ and $0.21$ for charmed and beauty meson decays, respectively. These correspond to the four-dimensional $\overline{\rm MS}$ masses at the scales of $c$ and $b$ quark masses, obtained through the running \cite{Chetyrkin:2000yt} of the PDG value, $m_s(2~\mathrm{GeV})=93~\mathrm{MeV}$ \cite{ParticleDataGroup:2020ssz}. We vary heavy quark masses with some reference values taken from the PDG data, while we set $m_u=m_d=0$. In the $B$ meson decays, charm quark mass is fixed by $m_c/\beta=4.9$, corresponding to the pole mass in four-dimensions.
\par
\begin{figure}[H]
 \begin{center}
 \includegraphics[width=140mm]{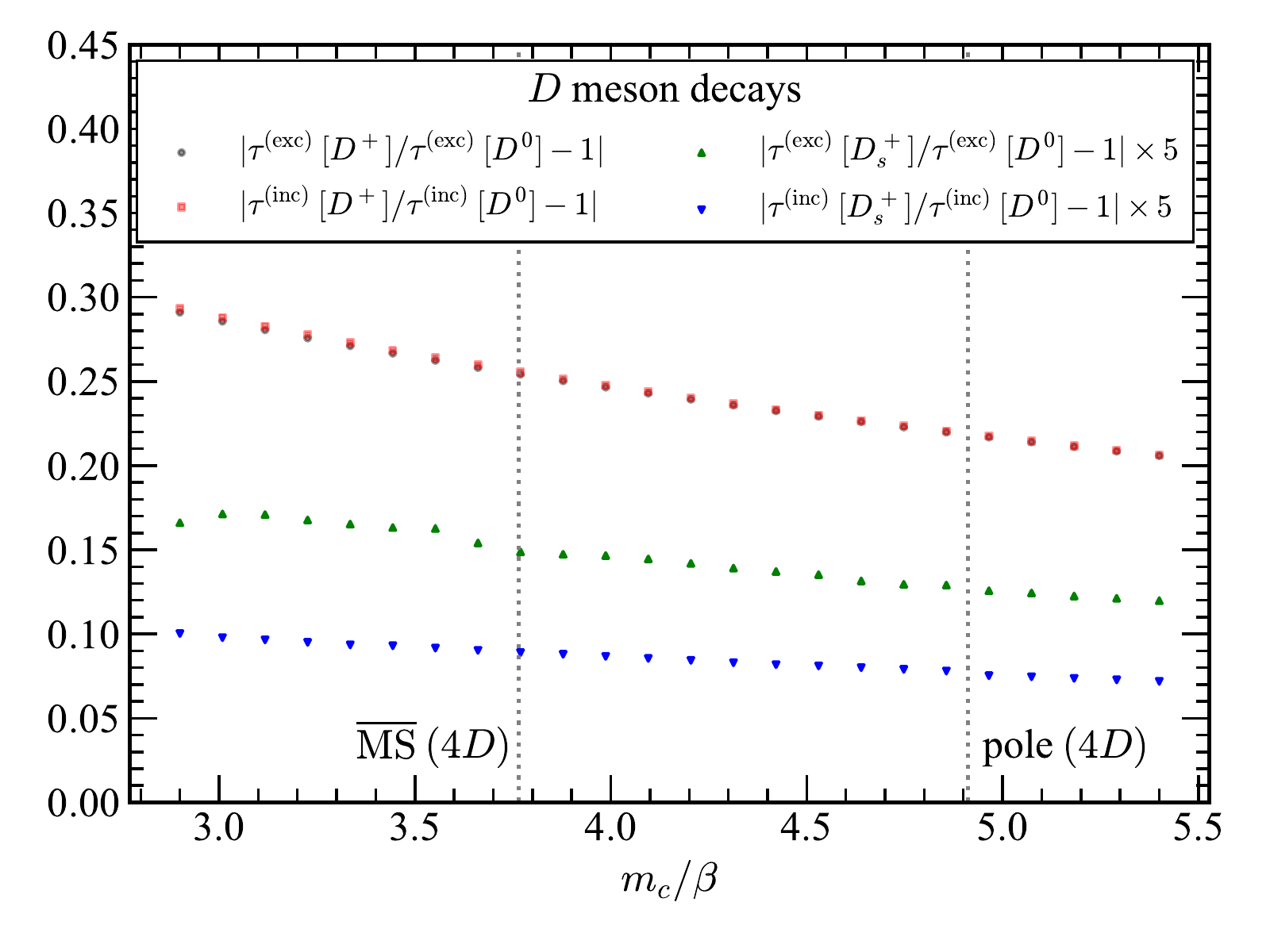}
 \end{center}
 \vspace{-8mm}
 \caption{Charm quark mass dependences of the absolute values for the deviation of the $D$ meson lifetime ratios from unity. The black dot and red square represent the cases of $D^+$ and $D^0$ lifetimes while the green triangle and blue inverse triangle denote the cases of $D_s^+$ and $D^0$ for exclusive and inclusive rates, respectively. Two dotted grey vertical lines correspond to the $\overline{\rm MS}$ mass at the scale of charm quark mass and the pole mass from the left, both of which are defined in four-dimensions.}
 \label{Fig:2}
\end{figure}
\begin{figure}[H]
 \begin{center}
 \includegraphics[width=140mm]{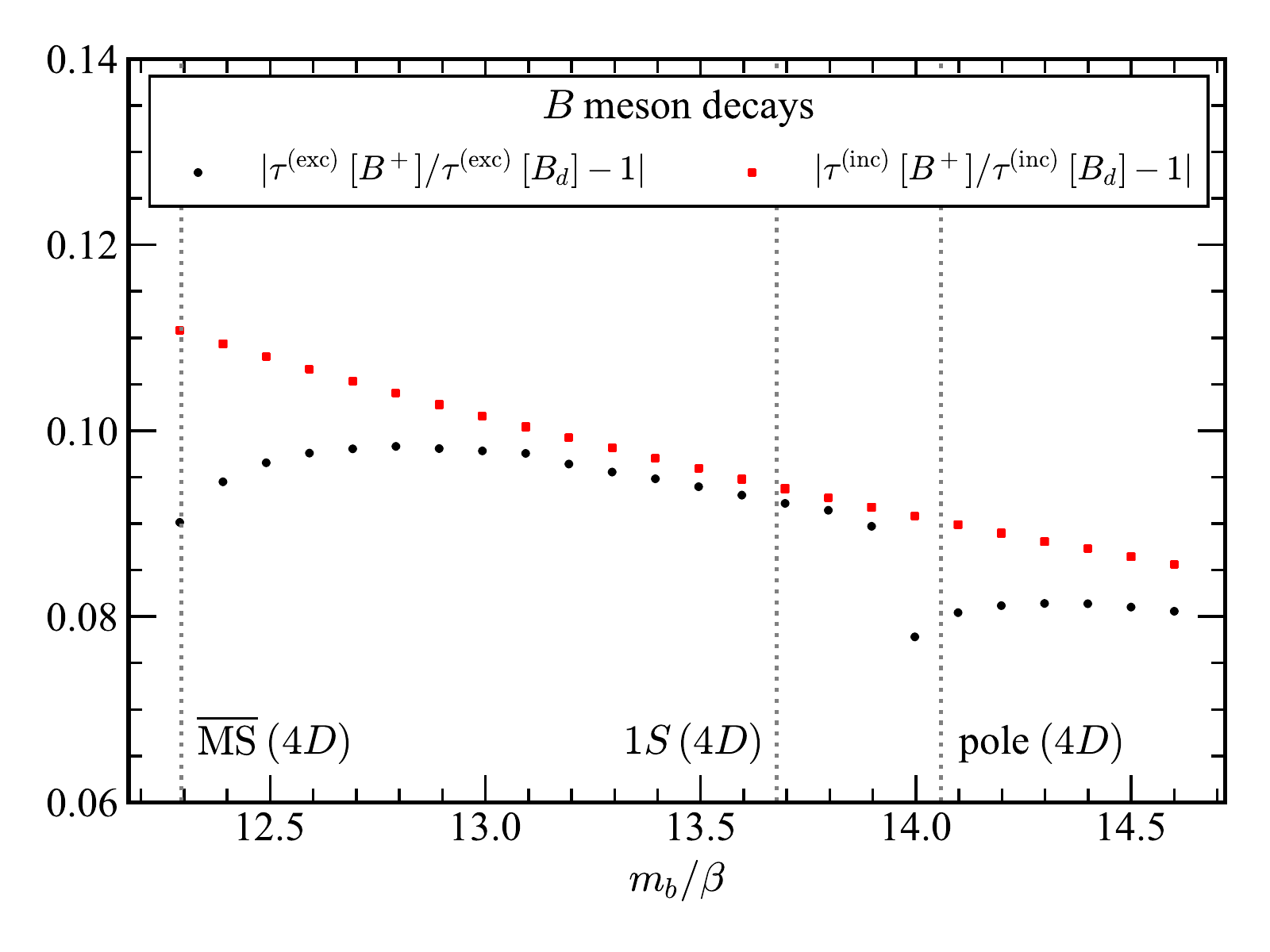}
 \end{center}
 \vspace{-8mm}
 \caption{Plot similar to Fig.~2 for beauty mesons. Three dotted grey vertical lines are $\overline{\rm MS}$ at the scale of bottom quark mass, the $1S$ scheme mass, and the pole mass from the left, all of which are defined in four-dimensions.}
 \label{Fig:3}
\end{figure}

The numerical results for decays of charmed and beauty mesons are given in Fig.~\ref{Fig:2} and Fig.~\ref{Fig:3}, respectively. The absolute values for the deviations of lifetime ratios from unity, whose signs are fixed to positive regardless of the spacetime dimensions, are plotted in those figures. The numerical stabilities are confirmed by varying the number of points for the eigenvectors by 50 units. In both Figs.~\ref{Fig:2} and Fig.~\ref{Fig:3}, one can verify that duality in the sense of the agreement between exclusive and inclusive objects improves as the heavy quark mass gets heavier.
\par
For the $D$ meson decays in Fig.~\ref{Fig:2}, a first point to be noted is that $0.3>|\tau[D^+]/\tau[D^0]-1|>0.2$ is seen in the plotted domain. It is equivalent to $0.8>\tau[D^+]/\tau[D^0]>0.7$ as PI contributes constructively to the decay rate of $D^+$. This is to be contrasted with the four-dimensional case, $\tau[D^+]/\tau[D^0]-1\approx 1.5$. As was already mentioned, in two-dimensions the phase space enhancement of $16\pi^2$ is absent while PI gives $\mathcal{O}(1/m_Q)$ corrections, unlike the four-dimensional case. The former gives a suppression while the latter provides an enhancement. The net result in Fig.~\ref{Fig:2} shows that $|\tau[D^+]/\tau[D^0]-1|$ is suppressed relative to the four-dimensional case. Another point is that the difference between inclusive and exclusive results of $|\tau[D^+]/\tau[D^0]-1|$ is quite small, indicating excellent realization of duality. As for $|\tau[D^+_s]/\tau[D^0]-1|$, the difference between exclusive and inclusive objects is typically $\approx 0.02$  for $m_c^{\mathrm{pole} (4D)}\gtrsim m_c \gtrsim m_c^{\overline{\rm MS} (4D)}$, which is larger than the case of $|\tau[D^+]/\tau[D^0]-1|$, in the currently considered numerical accuracy.\par
For the beauty meson decays in Fig.~\ref{Fig:3}, the hadronic thresholds for exclusive processes are more obviously seen than the charmed meson decays in Fig.~\ref{Fig:2}. In this case, it is shown that agreement between inclusive and exclusive $\tau[B^+]/\tau[B^0]$ is not as good as the case for $\tau[D^+]/\tau[D^0]$.
\par
In Tab.~1, the maximal difference between inclusive and exclusive lifetime ratios in the range of $m_Q^{\mathrm{pole} (4D)}\gtrsim m_Q \gtrsim m_Q^{\overline{\rm MS} (4D)}$ $(Q=c, b)$ is compared to the current uncertainties of the four-dimensional results. For the lifetime ratios of charmed mesons, the size of duality violation obtained in this work is smaller than the theoretical uncertainties in $4D$. Furthermore, we find that duality violation for $\tau[D^+]/\tau[D^0]$ $(\tau[D^+_s]/\tau[D^0])$ is smaller than (comparable with) the current experimental uncertainty. As to beauty mesons, the size of duality violation in $\tau[B^+]/\tau[B^0]$ is comparable with the current uncertainty of the HQE and is much larger than the experimental error. However, it should be noted that PI is not enhanced by 16$\pi^2$ and  and gives the $\mathcal{O}(1/m_Q)$ effect so that in comparing the result with one in 4D, overall scaling for $|\tau[D^+]/\tau[D^0]-1|$ and $|\tau[B^+]/\tau[B^0]-1|$ in 2D is somewhat subtle.
%===================
%===================
\begin{table}[ht]
\caption{Comparison between duality violation in 2D and the theoretical and experimental results in 4D for the lifetime ratios. The second column shows the maximal deviation between inclusive and exclusive lifetime ratios for $m_Q^{\rm pole, 4D}\gtrsim m_Q\gtrsim m_Q^{\overline{\rm MS}, 4D}$ $(Q=c, b)$ in 2D. Third and fourth column represent the recent HQE results and the experimental values, respectively, both in 4D. For the lifetime of $D_s^+$, the contribution of $D_s^+\to \tau^+\nu_\tau$ is properly subtracted \cite{Lenz:2013aua, Cheng:2018rkz, King:2021xqp}.}
\centering 
\begin{tabular}{l l l l} 
\hline\hline
$\quad$ Mode & Difference b/w & $\qquad\quad$HQE ($4D$) & Exp. ($4D$) \cite{ParticleDataGroup:2020ssz} \rule[0mm]{0mm}{4mm}\\ 
$\quad$  & inc. and exc. ($2D$)& $\qquad\quad$&  \\ 
\hline 
$\tau[D^+]/\tau[D^0]$ & \qquad$\approx 0.001$ & $2.80\pm0.85^{+0.01+0.11}_{-0.14-0.26}$ \cite{King:2021xqp}& $2.54\pm0.02$ \rule[0mm]{0mm}{4mm}\\ 
$\tau[D^+_s]/\tau[D^0]$  & $\qquad\approx 0.02$ & $1.01\pm 0.15^{+0.02+0.01}_{-0.03-0.01}$ \cite{King:2021xqp} & $1.30\pm 0.01$ \\
$\tau[B^+]/\tau[B^0]$ & $\qquad\approx 0.02$ & $1.074^{+0.017}_{-0.016}$ \cite{Cheng:2018rkz}& $1.076\pm0.004$ \\
\hline 
\end{tabular}
\label{table:nonlin}
\end{table}
%=======================
\section{Conclusion}
%=======================
In this letter, we have studied quark-hadron duality for lifetime ratios in the 't Hooft model. The numerical results presented in this work supposedly gives a qualitative estimate of duality violation for color-allowed contributions to heavy meson decays. Further improvement in accuracy entails the finite-$N_c$ corrections from WA and also color-suppressed contributions as well as the inclusion of the triple overlap integral, which is a technical task.
\par
We have investigated the difference between inclusive and exclusive lifetimes by adopting four-dimensional quark masses as reference values. For the lifetime ratios characterized by PI, $\tau[D^+]/\tau[D^0]$ and $\tau[B^+]/\tau[B^0]$, duality is particularly good for the former, while the hadronic thresholds are more obviously seen for the latter. Duality violation for the former is smaller than both theoretical and experimental errors in the realistic 4D results. Meanwhile, to the certain accuracy of the numerical calculation, the maximal sizes of duality violation for $\tau[D^+_s]/\tau[D^0]$ and $\tau[B^+]/\tau[B^0]$ are comparable with the current experimental and theoretical uncertainties in 4D, respectively.  In order to reveal whether duality violation is negligible in reality, it is certainly encouraged that theoretical precision of the HQE is improved through the higher order perturbative QCD corrections as well as precision determination of the matrix elements.
%=======================
\section*{Acknowledgements}\noindent
The author would like to thank Hai-Yang Cheng for reading the manuscript and useful comments. This work was supported in part by MOST of R.O.C. under Grant No.~MOST-110-2811-M-001-540-MY3.
%=======================
%%%%%%%%%%%%
%%% References %%%
%%%%%%%%%%%%


\begin{thebibliography}{99}

%\cite{Wilson:1969zs}
\bibitem{Wilson:1969zs}
K.~G.~Wilson, ``Nonlagrangian models of current algebra,''
Phys. Rev. \textbf{179}, 1499-1512 (1969);
K.~G.~Wilson and J.~B.~Kogut,
``The Renormalization group and the epsilon expansion,''
Phys. Rept. \textbf{12}, 75-199 (1974).

%\cite{Shifman:1978bx}
\bibitem{Shifman:1978bx}
M.~A.~Shifman, A.~I.~Vainshtein and V.~I.~Zakharov,
``QCD and Resonance Physics. Theoretical Foundations,''
Nucl. Phys. B \textbf{147}, 385-447 (1979);
M.~A.~Shifman, A.~I.~Vainshtein and V.~I.~Zakharov,
``QCD and Resonance Physics: Applications,''
Nucl. Phys. B \textbf{147}, 448-518 (1979);
V.~A.~Novikov, M.~A.~Shifman, A.~I.~Vainshtein and V.~I.~Zakharov,
``Wilson's Operator Expansion: Can It Fail?,''
Nucl. Phys. B \textbf{249}, 445-471 (1985).

%\cite{Bigi:1997fj}
\bibitem{Bigi:1997fj}
I.~I.~Y.~Bigi, M.~A.~Shifman and N.~Uraltsev,
``Aspects of heavy quark theory,''
Ann. Rev. Nucl. Part. Sci. \textbf{47}, 591-661 (1997)
[arXiv:hep-ph/9703290 [hep-ph]].

%\cite{Lenz:2014jha}
\bibitem{Lenz:2014jha}
A.~Lenz,
``Lifetimes and heavy quark expansion,''
Int. J. Mod. Phys. A \textbf{30}, no.10, 1543005 (2015)
[arXiv:1405.3601 [hep-ph]].

%\cite{Kirk:2017juj}
\bibitem{Kirk:2017juj}
M.~Kirk, A.~Lenz and T.~Rauh,
``Dimension-six matrix elements for meson mixing and lifetimes from sum rules,''
JHEP \textbf{12}, 068 (2017)
[erratum: JHEP \textbf{06}, 162 (2020)]
[arXiv:1711.02100 [hep-ph]].

%\cite{Cheng:2018rkz}
\bibitem{Cheng:2018rkz}
H.~Y.~Cheng,
``Phenomenological Study of Heavy Hadron Lifetimes,''
JHEP \textbf{11}, 014 (2018)
[arXiv:1807.00916 [hep-ph]].

%\cite{King:2019lal}
\bibitem{King:2019lal}
D.~King, A.~Lenz and T.~Rauh,
``B$_{s}$ mixing observables and $|V_{td}/V_{ts}|$ from sum rules,''
JHEP \textbf{05}, 034 (2019)
[arXiv:1904.00940 [hep-ph]].

%\cite{HFLAV:2019otj}
\bibitem{HFLAV:2019otj}
Y.~S.~Amhis \textit{et al.} [HFLAV],
``Averages of b-hadron, c-hadron, and $\tau $-lepton properties as of 2018,''
Eur. Phys. J. C \textbf{81}, no.3, 226 (2021)
[arXiv:1909.12524 [hep-ex]].

%\cite{Gaillard:1974mw}
\bibitem{Gaillard:1974mw}
M.~K.~Gaillard, B.~W.~Lee and J.~L.~Rosner,
``Search for Charm,''
Rev. Mod. Phys. \textbf{47}, 277-310 (1975).

%\cite{Ellis:1975hr}
\bibitem{Ellis:1975hr}
J.~R.~Ellis, M.~K.~Gaillard and D.~V.~Nanopoulos,
``On the Weak Decays of High Mass Hadrons,''
Nucl. Phys. B \textbf{100}, 313 (1975)
[erratum: Nucl. Phys. B \textbf{104}, 547 (1976)].

%\cite{Cabibbo:1977zv}
\bibitem{Cabibbo:1977zv}
N.~Cabibbo and L.~Maiani,
``Two-Body Decays of Charmed Mesons,''
Phys. Lett. B \textbf{73}, 418 (1978)
[erratum: Phys. Lett. B \textbf{76}, 663 (1978)].

%\cite{Koide:1979iw}
\bibitem{Koide:1979iw}
Y.~Koide,
``Inclusive Studies of Charmed Meson Decays and SU(4) 20-plet Dominance Model,''
Phys. Rev. D \textbf{20}, 1739 (1979)
[erratum: Phys. Rev. D \textbf{21}, 853 (1980)].

%\cite{Sawayanagi:1982ms}
\bibitem{Sawayanagi:1982ms}
H.~Sawayanagi, K.~Fujii, T.~Okazaki and S.~Okubo,
``On the $D^0-D^+$ Lifetime Problem: Evaluation of Interference and $W$ Exchange Effects,''
Phys. Rev. D \textbf{27}, 2107 (1983).

%\cite{Belle-II:2021cxx}
\bibitem{Belle-II:2021cxx}
F.~Abudin\'en \textit{et al.} [Belle-II],
``Precise measurement of the $D^0$ and $D^+$ lifetimes at Belle II,''
[arXiv:2108.03216 [hep-ex]].

%\cite{Glashow:1970gm}
\bibitem{Glashow:1970gm}
S.~L.~Glashow, J.~Iliopoulos and L.~Maiani,
``Weak Interactions with Lepton-Hadron Symmetry,''
Phys. Rev. D \textbf{2}, 1285-1292 (1970).

%\cite{Golowich:2005pt}
\bibitem{Golowich:2005pt}
E.~Golowich and A.~A.~Petrov,
``Short distance analysis of $D^0 - \bar{D^0}$ mixing,''
Phys. Lett. B \textbf{625}, 53-62 (2005)
[arXiv:hep-ph/0506185 [hep-ph]].

%\cite{Bobrowski:2010xg}
\bibitem{Bobrowski:2010xg}
M.~Bobrowski, A.~Lenz, J.~Riedl and J.~Rohrwild,
``How Large Can the SM Contribution to CP Violation in $D^0-\bar D^0$ Mixing Be?,''
JHEP \textbf{03}, 009 (2010)
[arXiv:1002.4794 [hep-ph]].

%\cite{Li:2020xrz}
\bibitem{Li:2020xrz}
H.~N.~Li, H.~Umeeda, F.~Xu and F.~S.~Yu,
``$D$ meson mixing as an inverse problem,''
Phys. Lett. B \textbf{810}, 135802 (2020)
[arXiv:2001.04079 [hep-ph]].

%\cite{Bloom:1970xb}
\bibitem{Bloom:1970xb}
E.~D.~Bloom and F.~J.~Gilman,
``Scaling, Duality, and the Behavior of Resonances in Inelastic electron-Proton Scattering,''
Phys. Rev. Lett. \textbf{25}, 1140 (1970).

%\cite{Poggio:1975af}
\bibitem{Poggio:1975af}
E.~C.~Poggio, H.~R.~Quinn and S.~Weinberg,
``Smearing the Quark Model,''
Phys. Rev. D \textbf{13}, 1958 (1976).

%\cite{Shifman:2000jv}
\bibitem{Shifman:2000jv}
M.~A.~Shifman,
``Quark hadron duality,''
[arXiv:hep-ph/0009131 [hep-ph]].

%\cite{Chay:1994si}
\bibitem{Chay:1994si}
J.~Chay and S.~J.~Rey,
``Instanton contribution to $B \to X_u e \bar{\nu}$ decay,''
Z. Phys. C \textbf{68}, 431-438 (1995)
[arXiv:hep-ph/9404214 [hep-ph]];
``Instanton contribution to $B \to X_s \gamma$ decay,''
Z. Phys. C \textbf{68}, 425-430 (1995)
[arXiv:hep-ph/9406279 [hep-ph]].

%\cite{Falk:1995yc}
\bibitem{Falk:1995yc}
A.~F.~Falk and A.~Kyatkin,
``Instantons and the endpoint of the lepton energy spectrum in charmless semileptonic B decays,''
Phys. Rev. D \textbf{52}, 5049-5055 (1995)
[arXiv:hep-ph/9502248 [hep-ph]].

%\cite{Chibisov:1996wf}
\bibitem{Chibisov:1996wf}
B.~Chibisov, R.~D.~Dikeman, M.~A.~Shifman and N.~Uraltsev,
``Operator product expansion, heavy quarks, QCD duality and its violations,''
Int. J. Mod. Phys. A \textbf{12}, 2075-2133 (1997)
[arXiv:hep-ph/9605465 [hep-ph]].

%\cite{Shifman:1994yf}
\bibitem{Shifman:1994yf}
M.~A.~Shifman,
``Theory of preasymptotic effects in weak inclusive decays,''
[arXiv:hep-ph/9405246 [hep-ph]];
``Recent progress in the heavy quark theory,''
[arXiv:hep-ph/9505289 [hep-ph]].

%\cite{Zhitnitsky:1995qa}
\bibitem{Zhitnitsky:1995qa}
A.~R.~Zhitnitsky,
``Lessons from QCD in two-dimensions ($N \to \infty$): Vacuum structure, asymptotic series, instantons and all that...,''
Phys. Rev. D \textbf{53}, 5821-5833 (1996)
[arXiv:hep-ph/9510366 [hep-ph]].

%\cite{Colangelo:1997ni}
\bibitem{Colangelo:1997ni}
P.~Colangelo, C.~A.~Dominguez and G.~Nardulli,
``Violations of local duality in the heavy quark sector,''
Phys. Lett. B \textbf{409}, 417-424 (1997)
[arXiv:hep-ph/9705390 [hep-ph]].

%\cite{Grinstein:1997xk}
\bibitem{Grinstein:1997xk}
B.~Grinstein and R.~F.~Lebed,
``Explicit quark-hadron duality in heavy-light meson weak decays in the 't Hooft model,''
Phys. Rev. D \textbf{57}, 1366-1378 (1998)
[arXiv:hep-ph/9708396 [hep-ph]].

%\cite{Blok:1997hs}
\bibitem{Blok:1997hs}
B.~Blok, M.~A.~Shifman and D.~X.~Zhang,
``An Illustrative example of how quark hadron duality might work,''
Phys. Rev. D \textbf{57}, 2691-2700 (1998)
[erratum: Phys. Rev. D \textbf{59}, 019901 (1999)]
[arXiv:hep-ph/9709333 [hep-ph]].

%\cite{Bigi:1998kc}
\bibitem{Bigi:1998kc}
I.~I.~Y.~Bigi, M.~A.~Shifman, N.~Uraltsev and A.~I.~Vainshtein,
``Heavy flavor decays, OPE and duality in two-dimensional 't Hooft model,''
Phys. Rev. D \textbf{59}, 054011 (1999)
[arXiv:hep-ph/9805241 [hep-ph]].

%\cite{Grinstein:2}
\bibitem{Grinstein:2}
B.~Grinstein and R.~F.~Lebed,
``Quark hadron duality in the 't Hooft model for meson weak decays: Different quark diagram topologies,''
Phys. Rev. D \textbf{59}, 054022 (1999)
[arXiv:hep-ph/9805404 [hep-ph]].

%\cite{Bigi:1999fi}
\bibitem{Bigi:1999fi}
I.~I.~Y.~Bigi and N.~Uraltsev,
``Heavy quark expansion and preasymptotic corrections to decay widths in the 't Hooft model,''
Phys. Rev. D \textbf{60}, 114034 (1999)
[arXiv:hep-ph/9902315 [hep-ph]]; ``Pauli interference in the 't Hooft model: Heavy quark expansion and quark hadron duality,''
Phys. Lett. B \textbf{457}, 163-169 (1999)
[arXiv:hep-ph/9903258 [hep-ph]].

%\cite{Burkardt:2000ez}
\bibitem{Burkardt:2000ez}
M.~Burkardt and N.~Uraltsev,
``Analytical heavy quark expansion in the 't Hooft model,''
Phys. Rev. D \textbf{63}, 014004 (2001)
[arXiv:hep-ph/0005278 [hep-ph]].

%\cite{Lebed:1}
\bibitem{Lebed:1}
R.~F.~Lebed and N.~G.~Uraltsev,
``Precision studies of duality in the 't Hooft model,''
Phys. Rev. D \textbf{62}, 094011 (2000)
[arXiv:hep-ph/0006346 [hep-ph]].

%\cite{Beane:2001uj}
\bibitem{Beane:2001uj}
S.~R.~Beane,
``Constraining quark hadron duality at large $N_c$,''
Phys. Rev. D \textbf{64}, 116010 (2001)
[arXiv:hep-ph/0106022 [hep-ph]].

%\cite{Grinstein:2001zq}
\bibitem{Grinstein:2001zq}
B.~Grinstein,
``Global duality in heavy flavor decays in the 't Hooft model,''
Phys. Rev. D \textbf{64}, 094004 (2001)
[arXiv:hep-ph/0106205 [hep-ph]].

%\cite{Grinstein:2001nu}
\bibitem{Grinstein:2001nu}
B.~Grinstein,
``Global duality in heavy flavor hadronic decays,''
Phys. Lett. B \textbf{529}, 99-104 (2002)
[arXiv:hep-ph/0112323 [hep-ph]].

%\cite{Mondejar:2006ct}
\bibitem{Mondejar:2006ct}
J.~Mondejar, A.~Pineda and J.~Rojo,
``Heavy meson semileptonic differential decay rate in two dimensions in the large $N_c$,''
JHEP \textbf{09}, 060 (2006)
[arXiv:hep-ph/0605248 [hep-ph]].

%\cite{Mondejar:2008pi}
\bibitem{Mondejar:2008pi}
J.~Mondejar and A.~Pineda,
``Breakdown of the operator product expansion in the 't Hooft model,''
Phys. Rev. Lett. \textbf{101}, 152002 (2008)
[arXiv:0807.0011 [hep-ph]];
``Deep inelastic scattering and factorization in the 't Hooft Model,''
Phys. Rev. D \textbf{79}, 085011 (2009)
[arXiv:0901.3113 [hep-ph]].

%\cite{Umeeda:2021llf}
\bibitem{Umeeda:2021llf}
H.~Umeeda,
``Quark-hadron duality for heavy meson mixings in the \textquoteright{}t Hooft model,''
JHEP \textbf{09}, 066 (2021)
[arXiv:2106.06215 [hep-ph]].


%\cite{Blok:1996fg}
\bibitem{Blok:1996fg}
B.~Blok, M.~A.~Shifman and N.~Uraltsev,
``Chiral symmetry breaking and duality in the anti-Q q channel,''
Nucl. Phys. B \textbf{494}, 237-259 (1997)
[arXiv:hep-ph/9610515 [hep-ph]].

%\cite{Blok:1997yd}
\bibitem{Blok:1997yd}
B.~Blok and M.~Lublinsky,
``Parton-hadron duality in QCD sum rules: Quantum mechanical examples,''
Phys. Rev. D \textbf{57}, 2676-2690 (1998)
[erratum: Phys. Rev. D \textbf{58}, 019903 (1998)]
[arXiv:hep-ph/9706484 [hep-ph]].

%\cite{tHooft:1974pnl}
\bibitem{tHooft:1974pnl}
G.~'t Hooft,
``A Two-Dimensional Model for Mesons,''
Nucl. Phys. B \textbf{75}, 461-470 (1974).

%\cite{Lenz:2013aua}
\bibitem{Lenz:2013aua}
A.~Lenz and T.~Rauh,
``D-meson lifetimes within the heavy quark expansion,''
Phys. Rev. D \textbf{88}, 034004 (2013)
[arXiv:1305.3588 [hep-ph]].

%\cite{King:2021xqp}
\bibitem{King:2021xqp}
D.~King, A.~Lenz, M.~L.~Piscopo, T.~Rauh, A.~V.~Rusov and C.~Vlahos,
``Revisiting Inclusive Decay Widths of Charmed Mesons,''
[arXiv:2109.13219 [hep-ph]].

%\cite{Guberina:1979xw}
\bibitem{Guberina:1979xw}
B.~Guberina, S.~Nussinov, R.~D.~Peccei and R.~Ruckl,
``D-Meson Lifetimes and Decays,''
Phys. Lett. B \textbf{89}, 111-115 (1979).

%\cite{Nambu:1950dpa}
\bibitem{Nambu:1950dpa}
Y.~Nambu,
``Force potentials in quantum field theory,''
Prog. Theor. Phys. \textbf{5}, 614-633 (1950).

%\cite{Salpeter:1951sz}
\bibitem{Salpeter:1951sz}
E.~E.~Salpeter and H.~A.~Bethe,
``A Relativistic equation for bound state problems,''
Phys. Rev. \textbf{84}, 1232-1242 (1951).

%\cite{Buras:1998raa}
\bibitem{Buras:1998raa}
A.~J.~Buras,
``Weak Hamiltonian, CP violation and rare decays,''
[arXiv:hep-ph/9806471 [hep-ph]].

%\cite{Cabibbo:1963yz}
\bibitem{Cabibbo:1963yz}
N.~Cabibbo,
``Unitary Symmetry and Leptonic Decays,''
Phys. Rev. Lett. \textbf{10}, 531-533 (1963).

%\cite{Kobayashi:1973fv}
\bibitem{Kobayashi:1973fv}
M.~Kobayashi and T.~Maskawa,
``CP Violation in the Renormalizable Theory of Weak Interaction,''
Prog. Theor. Phys. \textbf{49}, 652-657 (1973).

%\cite{ParticleDataGroup:2020ssz}
\bibitem{ParticleDataGroup:2020ssz}
P.~A.~Zyla \textit{et al.} [Particle Data Group],
``Review of Particle Physics,''
PTEP \textbf{2020}, no.8, 083C01 (2020).

%\cite{Brower:1978wm}
\bibitem{Brower:1978wm}
R.~C.~Brower, W.~L.~Spence and J.~H.~Weis,
``Bound States and Asymptotic Limits for {QCD} in Two-dimensions,''
Phys. Rev. D \textbf{19}, 3024 (1979)

%\cite{Burkardt:2000uu}
\bibitem{Burkardt:2000uu}
M.~Burkardt,
``Off forward parton distributions in (1+1)-dimensional QCD,''
Phys. Rev. D \textbf{62}, 094003 (2000)
[arXiv:hep-ph/0005209 [hep-ph]].

%\cite{Jia:2017uul}
\bibitem{Jia:2017uul}
Y.~Jia, S.~Liang, L.~Li and X.~Xiong,
``Solving the Bars-Green equation for moving mesons in two-dimensional QCD,''
JHEP \textbf{11}, 151 (2017)
[arXiv:1708.09379 [hep-ph]].

%\cite{Chetyrkin:2000yt}
\bibitem{Chetyrkin:2000yt}
K.~G.~Chetyrkin, J.~H.~Kuhn and M.~Steinhauser,
``RunDec: A Mathematica package for running and decoupling of the strong coupling and quark masses,''
Comput. Phys. Commun. \textbf{133}, 43-65 (2000)
[arXiv:hep-ph/0004189 [hep-ph]].


\end{thebibliography}
\end{document}